\documentclass[preprint,prc,aps,tightenlines,floatfix,showpacs]{revtex4-1}

\usepackage{graphicx}
\usepackage{bm}
\usepackage{amsmath}
\usepackage{amssymb}

\begin{document}

\title{Structure of $^{16}$C from analyses of proton scattering data and the $B(E2)$ problem}

\author{S. Karataglidis$^{(1,2)}$}
\author{K. Murulane$^{(1)}$} 
\affiliation{$^{(1)}$ Department of Physics, University of Johannesburg, P.O. Box 524, Auckland Park, 2006, South Africa }
\affiliation{$^{(2)}$ School of Physics, University of Melbourne, Victoria, 3010, Australia}

\date{\today}             

\begin{abstract}
The $B(E2)$ value for the decay of the $2^+_1$ state of $^{16}$C to ground has been the subject of much
discussion. Analyses assuming a simple model of two neutrons coupled to a $^{14}$C core of the available data, which 
extend over an order of magnitude, gives reasonable agreement but with
an inclusion of a large effective charge for the neutrons. To assess this situation, a large scale $(2+4)\hbar\omega$ shell
model
calculation of $^{16}$C has been made from which the wave functions have been used to obtain the $B(E2)$ value.
As a check, comparison is made with available data on the spectrum of $^{16}$C and intermediate-energy elastic proton scattering.
\end{abstract}
\maketitle

The structure of the heavy carbon isotopes, above $^{14}$C, are of interest given the closed $0p$
neutron shell at $^{14}$C. One then has a minimum configuration of $2\hbar\omega$ character in the states
of these nuclei, given the population of neutrons in the $sd$ shell. The open $0p$ shell for the protons 
suggests significant mixing of additional $\hbar\omega$ components. All the carbon isotopes exhibit neutron skins,
given the neutron separation energies are relatively large (energies in MeV): 1.218, 4.251,
0.73, 4.188, and 3.3, for $^{15}$C, $^{16}$C, $^{17}$C, $^{18}$C, and $^{20}$C, respecitvely 
\cite{Aj91,Ti93,Ti95,Ti98}. There is an indication that $^{17}$C may exhibit a neutron halo, given that its
separation energy is quite low compared to the other isotopes. However, from analyses of inelastic proton
scattering data \cite{Ka08}, the ground state of $^{17}$C appears to have a neutron density distribution
consistent more with a skin than a halo.

Given that $^{17}$C may be described within a collective model as a neutron coupled to $^{16}$C
it is important to understand the structure of $^{16}$C. Little is known of the spectrum of
$^{16}$C: the ground state is $0^+;2$, the first excited state at 1.77~MeV is a $2^+$ state, and the
second excited state is at 3.03~MeV \cite{Ti93}. The second excited state is tentatively assigned
$0^+$. At higher energies there are only a cluster of three states at $\sim 4$~MeV and a state at 6.11~MeV.

Recent shell-model calculations \cite{Fu07,Co10,Yu12} considered the structure of $^{16}$C using
effective interactions derived from free nucleon-nucleon ($NN$) interactions. Fujii \textit{et al.} \cite{Fu07}
calculated the spectrum of $^{16}$C in a no-core shell model, incorporating all shells from the $0s$ to
the $0f1p$ shell. They sought to explain the $B(E2)$ value for the decay from the first excited state. They
described the low $B(E2)$ value by the inclusion of both an effective operator and an effective interaction. However,
as the authors state, the shell model used is within an incomplete space in energy, so the removal of
centre-of-mass spuriosity is not exact. They also conclude in their analysis that the $B(E2)$ value is
sensitive to the value of the effective neutron charge. With both corrections, they obtain a $B(E2)$ value of 0.82~$e^2$fm$^4$,
which agrees reasonably well with the stated experimental value of 0.63~$e^2$fm$^4$ \cite{Fu07}.
As the minimum configurations admitted in the even-mass carbon isotopes heavier than $^{14}$C is $2\hbar\omega$, 
the space used is incomplete, as $2\hbar\omega$ components in $^{16}$C must necessarily include the $1p1h$
excitations from the $0d1s$  to the $0g1d2s$ shell, effective charges must be used to calculate electromagnetic
observables to account for the limitations of the assumed model. 

Measurements of the lifetime of the $2^+_1$ state \cite{Wi08,On09} in $^{16}$C suggest larger values of the
$B(E2)$. Wiedeking \textit{et al.} report a value of $4.15 \pm 0.73$~$e^2$fm$^4$ \cite{Wi08} obtained from a 
lifetime measurement of the $2^+_1$ state in $^{16}$C from
the $^9$Be($^9$Be,$2p$) fusion-evaporation reaction. The subsequent measurements of the lifetime by
Ong \textit{et al.} \cite{On09} report values for the $B(E2)$ from $1.4 \pm 0.6 \pm 0.4$ to
$2.7 \pm 0.2 \pm 0.7$~$e^2$fm$^4$, a large variation, but all a factor of two below that reported by
Wiedeking \textit{et al.} Ong \textit{et al.} attribute this reduction to including the $\gamma$-ray angular distribution into
the previous measurement, which leads to a reduction in the observed lifetime by a factor of four.

Guiding the analyses of the $B(E2)$ value in $^{16}$C has been the assumption that the ground state of $^{16}$C may
be described by a dominant configuration of $\nu (sd)^2$ coupled to a $^{14}$C core. This has been assumed by
Wiedeking \textit{et al.} \cite{Wi08} and in the shell model calculation of Corragio \textit{et al.} \cite{Co10}. Extensions to that model suggest the inclusion
of proton configurations would influence the $B(E2)$ \cite{Ma14},  while the inclusion of more complicated
neutron $sd$ shell configurations may also explain it \cite{Fo16}. In all cases, an effective neutron charge of $\sim 0.4e$
has been required in order to fit the measured/adopted value. 

As an extension beyond these simple models, we have performed a no-core
$(2+4)\hbar\omega$ shell model calculation for the positive parity states of $^{16}$C, using a single particle
basis encompassing the six major shells from the $0s_{1/2}$ to the $0h1f2p$ shells.  The model space
is complete in $2\hbar\omega$ while the only limitation in $4\hbar\omega$ components
is the exclusion of the (neutron) $1p1h$ components from the $0d1s$ to the $0i1g2d3s$ shell. The shell-model interaction
of Zheng \textit{et al.} \cite{Zh95} was used and the calculations performed using the code OXBASH \cite{Ox86}. We have also performed a complete $(0+2+4)\hbar\omega$ calculation of the ground state of $^{14}$C to test the assumption of a
$\nu(sd)^2$ model for $^{16}$C. The wave functions obtained for the ground states of both nuclei are
\begin{align}
\left| ^{14}\text{C}_{\text{gs}} \right\rangle & = 62.54\% \left| 0\hbar\omega \right\rangle + 21.03\%
\left| 2\hbar\omega \right\rangle + 16.43\% \left| 4\hbar\omega \right\rangle \nonumber \\
\left| ^{16}\text{C}_{\text{gs}} \right\rangle & = \hspace*{2.82cm} 72.82\% \left| 2\hbar\omega \right\rangle + 27.18\% \left| 4\hbar\omega 
\right\rangle.
\end{align}
While there are dominant components corresponding to the configuration $\nu(sd)^2$ coupled to the ground
state of $^{14}$C, the significant admixing of $4\hbar\omega$ components in the ground state of $^{16}$C suggests
a more complicated wave function.  Those components that include the $\nu(sd)^2$ constitute $\sim 60\%$ of
the total wave function: 22.63\% comes from $\nu(0d_{\frac{5}{2}})^2$ while 22.37\% is $\nu(1s_{\frac{1}{2}})^2$.
The other 40\% of the total wave function comes from more complicated configurations, including those involving
proton admixing.

\begin{figure}
\centering\scalebox{0.6}{\includegraphics*{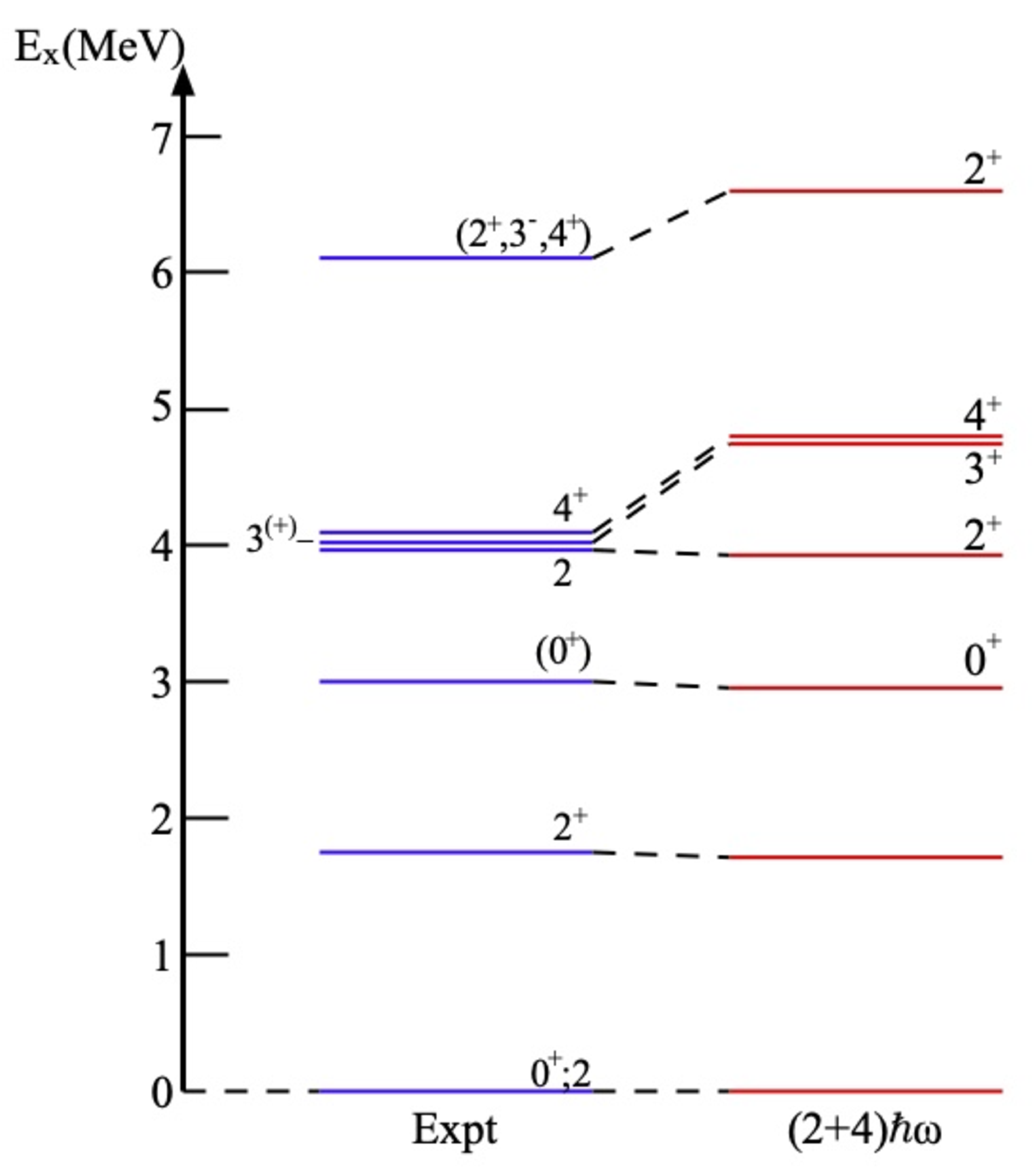}}
\caption{\label{c16spec} (Color online.) Low-energy spectrum of $^{16}$C. The experimental spectrum \cite{Ti93} is
compared to the result obtained from the $(2+4)\hbar\omega$ model described in text.}
\end{figure}

The full low-energy spectrum for $^{16}$C is shown in Fig.~\ref{c16spec}.
Therein, there is excellent agreement with the observed $2^+_1$ and $0^+_2$ states, while the cluster at 4~MeV is
reproduced reasonably well, with only a small separation between the calculated energies of the $2^+_2$, the $3^+$ and
$4^+$ states. The state observed at 6.11~MeV \cite{Ti93} is indicated as a third $2^+$ state in the model.

\begin{figure}
\centering\scalebox{0.6}{\includegraphics*{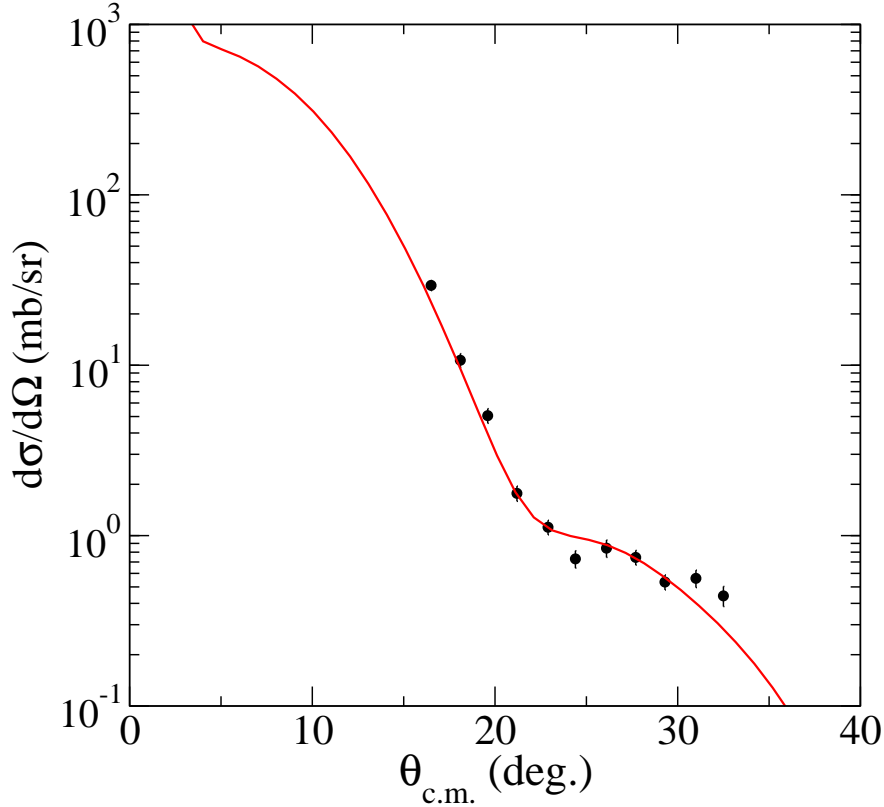}}
\caption{\label{c16scat} (Color online.) Differential cross section for the elastic scattering of 300~MeV protons from $^{16}$C.
The data \cite{Te14}, obtained in the inverse kinematics, are compared to the result of the microscopic optical
model calculation described in text.}
\end{figure}

To assess the wave functions obtained from the shell model, we have analysed the available data \cite{Te14}
the elastic scattering of 300~MeV protons from $^{16}$C. The microscopic Melbourne $g$-folding model for intermediate
energy nucleon-nucleus scattering \cite{Am00} was used, wherein the one-body density
matrix elements obtained from the ground state wave function was folded with the Bonn B nucleon-nucleon
interaction \cite{Ma89} to obtain the complex and nonlocal optical potential. The upper energy limit for the applicability
of the Melbourne $g$-folding model is 300~MeV \cite{Am00,De05}. Harmonic oscillators were assumed for the
single particle states in the nucleus, with oscillator parameter $b = 1.7$~fm which is appropriate for mass-16 nuclei
\cite{Ka96}. The result of the calculation for the differential
cross section so obtained is compared to the data \cite{Te14} in Fig.~\ref{c16scat}.
As shown in Fig.~\ref{c16scat}, the result of the $g$-folding calculation agrees very well with the data, with no fitting
required.

Given the large variation in the quoted $B(E2)$ values for the transition in $^{16}$C, as listed in Table~\ref{c16be2},
\begin{table}
\begin{ruledtabular}
\caption{\label{c16be2} Evaluated $B(E2)$ values for the transition $2^+_1 \rightarrow 0^+_1$ in $^{16}$C,
in units of $e^2$fm$^4$.}
\begin{tabular}{ll}
Author & $B(E2)$ value \\
\hline
Fujii \textit{et al.} \cite{Fu07} & $0.63 \pm 0.11 \pm 0.16$ \\
Wiedeking \textit{et al.} \cite{Wi08} & $4.15 \pm 0.73$ \\
Ong \textit{et al.} \cite{On09} & $2.7 \pm 0.2 \pm 0.7^a$ \\
& $2.4 \pm 0.4 \pm 0.6^b$ \\
& $1.4 \pm 0.6 \pm 0.4^c$ \\
Fortune \cite{Fo16} & $ 3.5 \pm 0.3^d$ 
\end{tabular}
\end{ruledtabular}
\begin{flushleft}
{\footnotesize
$^a$ Inelastic channel at $72A$~MeV.\\
$^b$ Breakup channel at $79A$~MeV. \\
$^c$ Inelastic channel at $40A$~MeV. \\
$^d$ Simple average of available data.
}
\end{flushleft}
\end{table}
we adopt the value given by Ong \textit{et al.} , 2.7~$e^2$fm$^4$, as the benchmark, and as that lying in the
middle of the range of values. From our shell model calculation, we find a $B(E2)$ value of 1.35~$e^2$fm$^4$,
using bare operators, well within the range of values indicated in Table~\ref{c16be2}. The inclusion of an effective charge of 
$0.09e$ gives a value of 2.79~$e^2$fm$^4$, while
one of $0.12e$ gives a value of 3.39~$e^2$fm$^4$.  These are much smaller values of effective charge than
those assumed in previous analyses. Together with the agreements found between experiment and model results for the spectrum
and the scattering, this indicates the large scale shell model adopted gives a far more reliable indication of the structure
of $^{16}$C. It is clear that the assumption of $\nu(sd)^2$ for the structure of $^{16}$C is too simplistic.

A large scale shell model calculation, in a $(2+4)\hbar\omega$ mode space, has been used to obtained the spectrum
and wave functions of $^{16}$C. There is very good agreement found between the results of the calculation for the
spectrum, 300~MeV elastic proton scattering, and the $B(E2)$ value, with experiment. This is especially so given
that has been no fitting to the data being described, except in the case of the $B(E2)$ value. In the latter case,
the bare operators give an acceptable value, within the range of the experimental values. The inclusion of
a much smaller effective charge than previously reported gives a value close to the somewhat larger values now
accepted. Overall, this suggests that the shell model calculations presented provides a far more reliable description
of the structures of $^{16}$C. The assumption of a simple $\nu(sd)^2$ structure, while indicated as a dominant component
of the total ground state wave function of $^{16}$C, is not entirely valid.
\bibliography{c16_papers}
\end{document}